\documentclass[%
 reprint,
superscriptaddress,
 amsmath,amssymb,
 aps,
pra,
]{revtex4-2}

\usepackage{amsthm,latexsym,amssymb,amsmath,verbatim}
\usepackage{textcomp}
\usepackage{gensymb}
\usepackage{graphicx}
\usepackage{dcolumn}
\usepackage{hyperref}
\usepackage{chemformula}
\usepackage[T1]{fontenc}
\usepackage{longtable}
\usepackage{multirow}
 \usepackage{booktabs}

\newcommand*{\fig}[1]{Figure~\ref{fig:#1}}

\begin{document}

\title{A common  origin of photoplastic and electroplastic effects in ZnS}

\author{Alexandra Fonseca Montenegro} 
\affiliation
{Department of Materials Science and Engineering, The Ohio State University, Columbus OH 43210, USA} 
\author{Sevim Genlik Polat} 
\affiliation
{Department of Materials Science and Engineering, The Ohio State University, Columbus OH 43210, USA} 
\author{Md Mohsinur Rahman Adnan}
\affiliation
{Department of Electrical and Computer Engineering, The Ohio State University, Columbus OH 43210, USA}
\author{Maryam Ghazisaeidi} 
\email{ghazisaeidi.1@osu.edu}
\affiliation
{Department of Materials Science and Engineering, The Ohio State University, Columbus OH 43210, USA} 
\affiliation
{Department of Physics, The Ohio State University, Columbus OH 43210, USA} 
\author{Roberto C. Myers} 
\email{myers.1079@osu.edu} 
\affiliation
{Department of Materials Science and Engineering, The Ohio State University, Columbus OH 43210, USA} 
\affiliation
{Department of Electrical and Computer Engineering, The Ohio State University, Columbus OH 43210, USA} 
\affiliation
{Department of Physics, The Ohio State University, Columbus OH 43210, USA} 

\begin{abstract}
Dislocation motion—the atomic-scale mechanism of crystal plasticity—governs the strength and ductility of materials. In functional materials, external stimuli beyond mechanical stress can also affect dislocation glide. In the wide band gap semiconductor ZnS, optical illumination suppresses plasticity, whereas electric fields can enhance dislocation motion. Here, we show that the common underlying mechanism for these phenomena is the charged dislocations that respond to the changes in  carrier concentration. Our prior theoretical work showed that locally charged dislocations in ZnS trap excess carriers, triggering core reconstructions that modify their mobility, with the positively charged Zn-rich core dislocations showing the most drastic change. Here, we validate this prediction experimentally by showing that either optical excitation or electronic doping selectively inhibits the glide of Zn-rich dislocations in epitaxially grown ZnS. First, imaging individual interface misfit dislocations under different optical excitation conditions shows that Zn-core glide is strongly reduced as optical power is increased, while the S-core dislocations show negligible sensitivity to light, marking the first, single misfit dislocation imaging of the photoplastic effect. Next, we show that a similar behavior is observed with direct electron (n-type) doping of ZnS epitaxial layers grown beyond the critical thickness. As the n-type dopant density is increased, the resulting Zn-core dislocation density is reduced by more than one order of magnitude, while the S-core density remains essentially unchanged, causing a sign reversal of the strain-anisotropy with n-type doping. These results demonstrate a common origin for the opto-electronic sensitivity of dislocations in ZnS and provide a pathway for the engineering of dislocation content in compound semiconductors.
\end{abstract}

\maketitle

\section{Introduction}\label{Intro}
\begin{figure*}
    \centering
    \includegraphics[width=\textwidth]{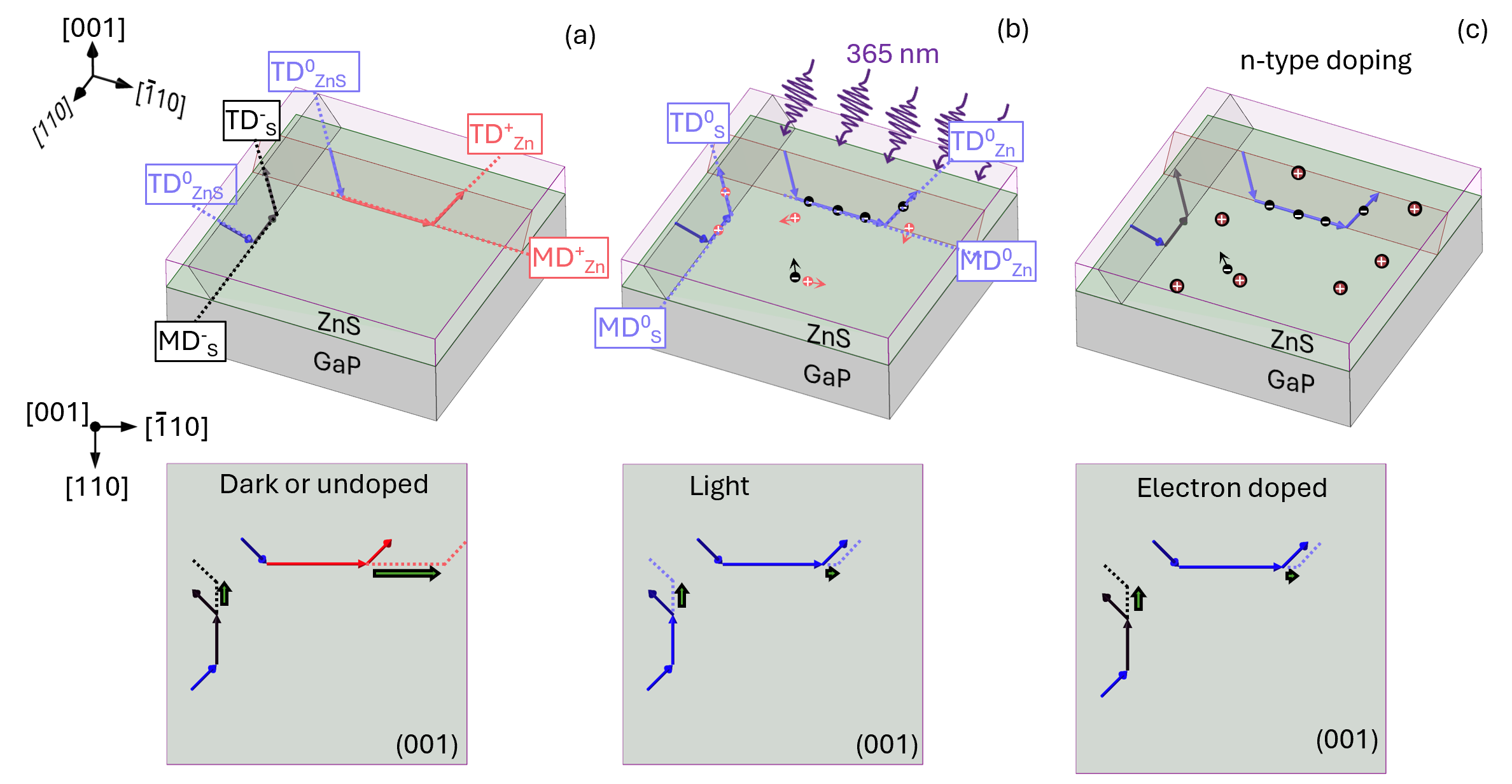}
    \caption{Charge-carrier controlled plasticity in ZnS epitaxial layers. (Top row) Perspective view and (bottom row) corresponding plan-view schematic of dislocation glide in ZnS epitaxially grown on GaP (001). Dashed lines in the bottom panels illustrate how misfit dislocation (MD) segments along the epitaxial interface lengthen by glide (green arrows) of surface terminating threading dislocation (TD) segments, whose (a) charge states and resulting glide mobilities are altered by (b) light or (c) electronic impurity doping. Each MD segment and its TD pair define a dislocation half-loop with a common \{111\} plane and burgers vector. (a) The core structure of an MD and one of its TDs ($\theta =60$°) are identical, having the edge-type characteristic associated with an extra half-plane terminating at an ionic core, the composition of which is locked to the MD line direction $\vec{u}$ due to the $\bar{4}$ symmetry of the crystal structure. In ZnS, the $\vec{u}$ = [110] MDs and one of its TDs are S$^-$-core ($MD^-_{S}$ and $TD^-_{S}$) and for $\vec{u}$ = [$\bar{1}$10], they are Zn$^+$-core ($MD^+_{Zn}$ and $TD^+_{Zn}$). The other TD segment is screw-type ($\theta =180$°) with a neutral ZnS composition ($TD^0_{ZnS}$). (b) 365 nm optical excitation generates photo-electrons (-) and holes (+) that are trapped by the Zn$^+$ and S$^-$ dislocations, respectively. (c) Al-doped ZnS leads to conduction band electrons (-) that get trapped at Zn$^+$ MDs and TDs leaving bound Al$^+$ donor-ions (+) in the bulk ZnS. Electrons trapped in Zn$^+$-core trigger new core reconstructions ($MD^0_{Zn}$ and $TD^0_{Zn}$) decreasing the glide mobility.}
    \label{fig:photoplastic}
\end{figure*}

Dislocations, topological one-dimensional defects, occur in all crystalline solids. These defects are one of the major deformation mechanisms in crystals, and their creation and motion under applied stress govern mechanical properties of solids such as strength and ductility. However, dislocations in electronic materials are thought to cause non-radiative recombination of charge carriers that leads to overheating of the crystal and malfunction of the device. Therefore, inhibiting the formation of these extended line defects is a central goal. Although researchers in structural materials and solid-state electronics pursue divergent goals when it comes to dislocations, engineering the motion of dislocations is a common goal; in the former, inhibiting or enhancing dislocation glide balances the trade-off between strength and ductility, while in the latter, spatially isolating dislocations from active device regions is the goal of solid-state synthesis and processing. In functionally active materials,  dislocation motion can be altered by external stimuli other than mechanical stress. A prominent example is the photoplastic effect in  ZnS where the material undergoes a brittle to ductile transition under different light conditions~\cite{OshimaYu2018}. In compound semiconductors, photo-exposure \cite{carlsson_increase_1969,carlsson_orientation_1971,garosshen_influence_1990} electron injection\cite{maeda_quantitative_1983}, and applied electric field\cite{LiMengqiang2023} can affect the flow stress (plasticity). However, the connection between the optical and electronic sensitivity of dislocations in ZnS is not yet well established.  Here we provide experimental evidence that the change in the charge carrier concentration is the common underlying mechanism for influencing the dislocation motion in ZnS.
In polar II-VI semiconductors, like ZnS, edge or mixed dislocations have either cation- or anion-rich cores and therefore are charged. Our previous first-principles calculations of dislocations in ZnS confirmed this.\cite{genlik_origin_2025} We also studied the effect of light, via photo-excited charge carriers on the dislocation core structure and mobility, and predicted that photoplasticity is caused by photo-electrons being trapped on the positively charged Zn-core dislocations, inducing a structural reconstruction within the dislocation core. As such, the energetic barrier for the dislocation to glide, the Peierls barrier, increased significantly for electron-doped Zn-core dislocations while the effect was minimal on the hole-doped S-cores. To test this prediction we study the effect of optical excitation as well as electronic doping on the formation of strain-relieving misfit dislocations (MDs) in epitaxial ZnS grown on GaP, shown schematically in Fig. \ref{fig:photoplastic}.

Networks of interface MDs can be controllably formed at closely lattice-matched film/substrate interfaces, e.g. ZnS/GaP (001), using layer-by-layer epitaxy.\cite{FonsecaAlexandra2024} For these strained epilayers, the combined restriction of the substrate orientation, tensile strain, symmetry of the zinc blende crystal structure, and slip system of ZnS force a strict relationship between the dislocation line direction $\vec{u}$, burgers vector $\vec{b}$, and slip plane (hkl), such that the core-structure of the dislocations can be inferred from $\vec{u}$ (Sec. \ref{dlngeometry}). 
Utilizing this inherent core structure anisotropy, we target the glide of Zn-core vs S-core dislocations. First, we show that our approach is capable of detailed characterization of the dislocation glide under optical excitation, presenting the first single-dislocation imaging of the photoplastic effect (Sec. \ref{single-photoplastic}).  Next, we study the effect of carrier concentration by direct n-type doping of ZnS epitaxial layers, grown beyond the critical thickness, with Al substitutional impurities (Sec. \ref{doping-plasticity}). As the dopant density is increased, the resulting Zn-core dislocation density is reduced by more than one order of magnitude, while the S-core density remains essentially unchanged, causing a sign reversal of the strain-anisotropy with n-type doping.  These results demonstrate a common origin for the opto-electronic sensitivity of dislocations in ZnS via changes in the carrier concentration.

\section{Dislocation Geometry}\label{dlngeometry}
Before relaxation commences, the epilayers are biaxially strained due to the lattice mismatch, $f$ = 0.56\%, with no difference in orthogonal in-plane directions. Above a critical thickness ($\sim$15 nm), strain-relaxation proceeds by nucleation and growth of misfit dislocation (MD) segments along the heteroepitaxial (001) interface, as schematically shown in Fig. \ref{fig:photoplastic}(a). The MDs terminate at the surface of the crystal by vertical threading dislocation (TD) segments. MD line length continues to increase by glide of the TD segments, which provides epitaxial strain-relief.\cite{freund_driving_1990} As in other zinc-blende structure epitaxial systems,\cite{brown_anisotropic_1989,matragrano_anisotropic_1996} ZnS/GaP displays anisotropy in the TD glide velocity along the orthogonal [110] versus the [$\bar{1}$10] directions, leading to an asymmetry in the MD populations and the resulting strain-relaxation \cite{FonsecaAlexandra2024}. This anisotropic mobility results in longer MD segments along the [$\bar{1}$10] direction, as depicted in Fig. \ref{fig:photoplastic}(a).

The anisotropy of strain-relaxation in these cubic materials is caused by an underlying asymmetry in the atomistic pathway for dislocation nucleation and glide. For zincblende III-V and II-VI compounds, there is a  difference in local bonding between the core structures for dislocations along [110] versus [$\bar{1}$10]. In tensile strained epitaxial zinc blende (AB) systems, the strain-relieving component of MDs consists of an extra half-plane inserted within the strained-layer, which terminates with either the A or the B ions at the interface \cite{abrahams_likesign_1972}. The in-equivalence of the [110] and [$\bar{1}$10] directions in this crystal structure, combined with the specific strain state and orientation of the film, place geometric restrictions on the $\vec{u}$ and $\vec{b}$ that would provide strain relaxation. The strain relieving MDs consist of $\vec{b}$ = a/2<110> on \{111\} slip planes, with $\theta=cos^{-1}(\hat{u}\cdot\hat{b})=60$° In our specific case of tensile strained (001) oriented ZnS with glide-type 60° (or 120°) dislocations,\cite{petrenko_charged_1980, osipyan_properties_1986} the half-plane  must terminate in an anion (S) for $\vec{u}$ = $\pm$[110], and a cation (Zn) for $\vec{u}$ = $\pm$[$\bar{1}$10] MDs. See Sec. \ref{App-dln} for full details.

An additional relationship exists between the MD core structure and their TD segments since each MD formed by nucleation and glide of a dislocation surface half-loop, with the bottom segment of the loop forming the strain-relieving MD and the remaining two sides forming the TDs. For example, in Fig. \ref{fig:photoplastic}(a) the black MD segment has $\vec{u}$ = [$\bar{1}\bar{1}$0], $\vec{b}$ = a/2[101], with a ($\bar{1}$11) slip plane, defining a 120° dislocation with an extra (1$\bar{1}$2) half-plane necessarily terminating in a line of S-anions. In intrinsic, undoped ZnS under dark conditions, the S-rich stoichiometry leads to a (-) charge state (black, MD$^-_S$) along these dislocations. Since the TDs are from the same half-loop, they share the same plane and $\vec{b}$ as the MD and the $\vec{u}$ is continuous. This defines the $\vec{u}$ of the TDs, where one segment has $\vec{u}=[\bar{1}0\bar{1}]$, which is 180° from $\vec{b}$, (blue), whereas the other TD segment (black) has $\vec{u}=[0\bar{1}1]$ such that $\theta=60$°. The 180° TD is a pure left-hand-screw which is electrically neutral (blue, TD$^0_{ZnS}$) containing no net stoichiometric imbalance, while the 60° (black, TD$^-_S$) shares the same (-) S-rich character as its MD. A second example is shown for orthogonal MD $\vec{u}$ = [$\bar{1}$10] on a (111) slip plane in Fig. \ref{fig:photoplastic}(a), where the 120° MD-core contains Zn-cations (red, MD$^+_{Zn}$). The TDs again consist of one (blue, TD$^0_{ZnS}$) 180° left-hand-screw TD as well as a 60° TD (red, TD$^+_{Zn}$) with a Zn-core matching that of the MD. For each of the matching MD/TD pairs, the core structure identity stems from the identical edge character (sin(60°)=sin(120°), extra half-planes), whereas the neutral screw components are equal in magnitude but with opposite chirality (left or right handed).  For the two remaining slip plane cases (not shown), the same general relation holds where each MD segment has a matching  charged TD (60° or 120°) with the same composition as its MD (either MD$^+_{Zn}$/TD$^+_{Zn}$, or  MD$^-_S$/TD$^-_S$), while the other TD is screw-type (0° right-hand or 180° left-hand) with a (blue) neutral ZnS core structure (TD$^0_{ZnS}$). Excess charge carriers, for example through photo-excitation (Fig. \ref{fig:photoplastic}(b)) or impurity doping (Fig. \ref{fig:photoplastic}(c))
will be captured at the oppositely charged dislocations. The epitaxial ZnS films therefore enable capturing the correlation between optical and electronic stimuli with the dislocation glide mobility and its dependence on core composition (Zn vs S-core). Next, we quantify the change in the anisotropic glide of the TDs in the two orthogonal directions with different carrier concentrations. 

\section{Imaging single dislocation photoplasticity}\label{single-photoplastic}
\begin{figure*}
    \centering
    \includegraphics[width=\textwidth]{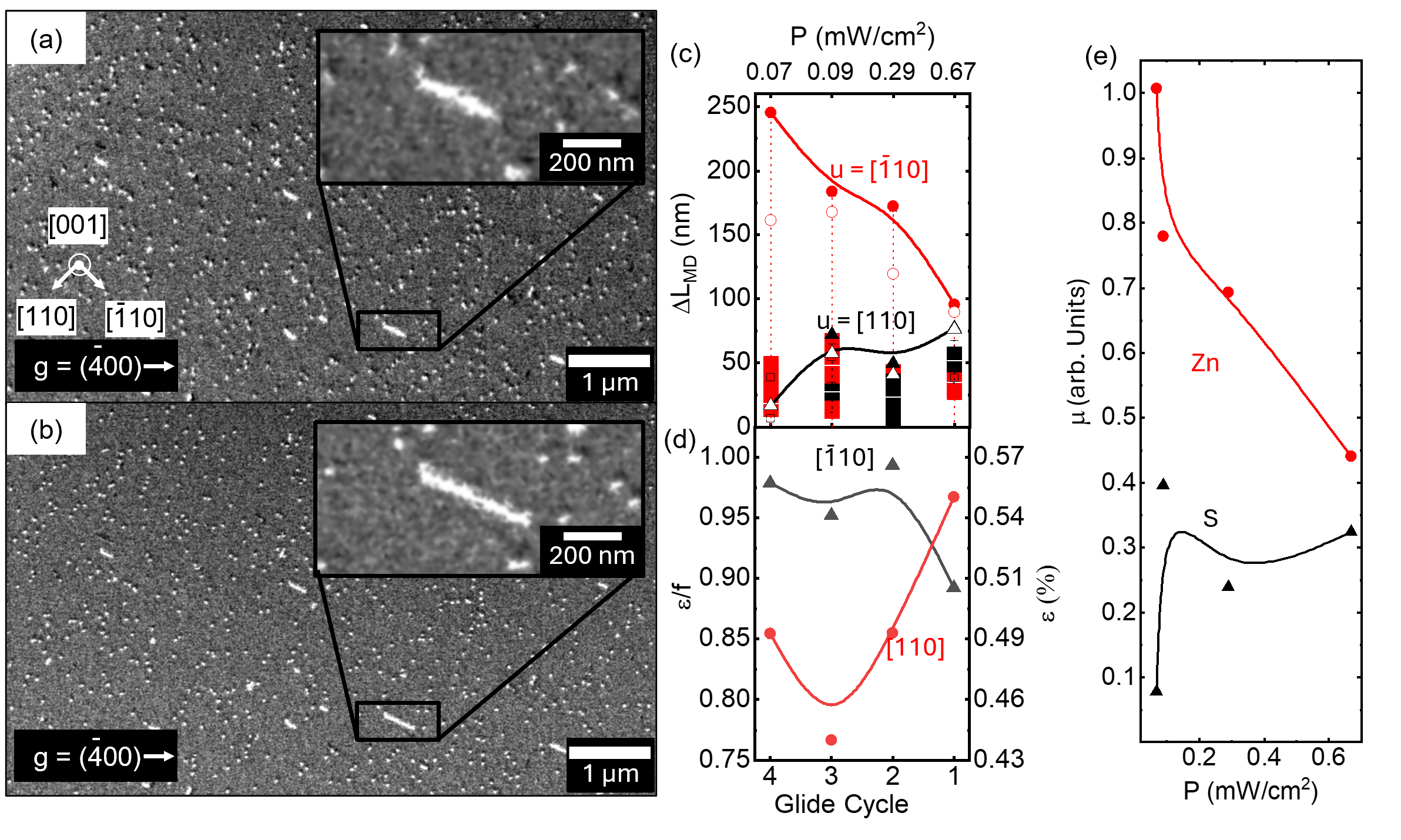}
  \caption{Single dislocation photoplastic effect from misfit dislocation (MD) imaging. Electron channeling contrast imaging (ECCI) of 20 nm undoped ZnS grown on GaP (001) (a) before and (b) after a glide cycle under 365 nm wavelength-irradiation at a power density, P = 0.07 mW/cm\textsuperscript{2}. Insets in (a) and (b) show the longest MD in the image from which the change in length after a glide cycle is measured, $\Delta L^{max}_{MD}$, plotted in (c) as solid symbols (with solid lines) in reverse experimental sequence for both the [$\bar{1}$10] Zn-core MDs and [110] S-core MDs. Top x-axis plots the P condition of each glide cycle. Open symbols are averages of the two max $\Delta L_{MD}$. Boxes (25\%-75\%) represent the full distribution of $\Delta L_{MD}$ measurements. (d) The average in-plane strain before and after each glide cycle $\epsilon$ (right axis) is calculated from the ECCI data and plotted for the two directions, as well as the strain-factor $\epsilon/f$ . The top x-axis shows P during each cycle. Normalizing $\Delta$ $L_{MD}^{max}$ data in (c) by the $\epsilon$ data in (d) yields the relative TD mobility $\mu$, which is plotted in (e) as a function of P for the two directions showing a strong reduction in $\mu_{Zn}$ and not $\mu_{S}$  with photoexcitation, see Fig. \ref{fig:photoplastic}(b). Lines are guides to the eye.}
    \label{fig:HeatExp}
\end{figure*}

While previous studies have explored the effect of light intensity on bulk ZnS \cite{OSHIMA2020690,ShenY2025}, here we utilize the well-defined dislocation populations contained in ZnS epilayers, to image the optical gating of individual dislocation glide. Scanning electron microscopy (SEM) based dislocation imaging via electron channeling contrast imaging (ECCI) is carried out on an undoped 20 nm thick ZnS epilayer grown by molecular beam epitaxy (MBE) on GaP (Sec. \ref{Method-MBE}). At this thickness, the low TD density ($\sim$6.5 $\mu m\textsuperscript{-2}$) enables the study of dislocation motion unhindered by interactions with neighboring dislocations, such that photo-induced changes in the glide mobility (photoplasticity) can be measured from imaging individual MDs. During the MBE growth, low densities of MD are generated by TD glide leading to a small amount of strain-relaxation. Given enough time or thermal energy, TDs would continue to glide until pinned by interaction with other dislocations. We use in-situ heating in the SEM to restart the strain-relaxation driven TD glide (Sec. \ref{Method-ECCI}). Representative ECCI images taken before and after a glide cycle are shown in Fig. \ref{fig:HeatExp}(a-b), illustrating an MD length change induced by glide of one of its TD segments.
For the dislocation identified in the inset, $\vec{u}$ = $\pm$[$\bar{1}$10], corresponding to a Zn-core MD. The change in the length of MDs before and after each glide cycle, $\Delta L_{MD}$, are measured under 365 nm incoherent continuous-wave illumination at different optical power densities (P). The results are plotted in Fig. \ref{fig:HeatExp}(c) in sequential order of the glide cycle (bottom x-axis), with the top x-axis labeling P. The majority of the MDs are grouped at small lengths due to pinning, which agrees with the log-normal distribution of MD lengths previously reported.\cite{FonsecaAlexandra2024} Thus, only the longest MDs in each distribution are most representative of the unimpeded TD glide velocity, which are plotted as solid symbols in Fig. \ref{fig:HeatExp}(c). It is clear that for $\vec{u}$ = $\pm$[$\bar{1}$10] (Zn-core) TD glide is inhibited by 365 nm light, while for the $\vec{u}$ = $\pm$[110] (S-core), TD glide perhaps shows a mild increase with 365 nm excitation.

In this experiment we track the change in MD length after a glide cycle, which is directly proportional to the average TD glide velocity, $v = \Delta L_{MD}^{max}/\Delta t$ where $\Delta t$ is the glide time. Assuming a simple dislocation mobility law,\cite{po_phenomenological_2016} $v=\tau b\mu = C\epsilon \mu$, where $\tau$ is the resolved shear stress,  $\mu$ is the average dislocation mobility, and $C=\tau b/\epsilon$ is a constant relating the $\tau$ to the in-plane strain $\epsilon$. Since the local strain state of the ZnS epilayer varies after each glide cycle due to relaxation induced from the previous cycle, as well as strain inhomogeneity from site to site, then the TDs in each cycle experience a different $\tau$, which would mask the underlying changes in $\mu$ expected due to the photoplastic effect. To account for this, we use quantitative ECCI data measurements of the linear MD density (total MD length per area) $\rho_{MD}$ to calculate the average $\epsilon$ before and after each glide cycle. The strain reduction $\delta$, along a direction orthogonal to an MD $\vec{u}$ is proportional to $\rho_{MD}$. Specifically, for 120° MDs with $\vec{b}$ = a/2<110>, along an (001) interface, \( \delta_{[110]}= f-\epsilon^{[110]}= (\sqrt{2}a/4)\rho_{MD}^{\bar{1}10} \), where a is the lattice constant of ZnS. The direction dependent $\epsilon$ are plotted in Fig. \ref{fig:HeatExp}(d). The results show as much as 20\% variation in the strain state with the largest change in the [110] direction. 

The direction-dependent MD measurements in Fig. \ref{fig:HeatExp}(c) are normalized by the respective strain driving force, to extract the relative TD mobility. Assuming $C\Delta t$ is constant from cycle to cycle, then $\mu_{Zn}=v^{[\bar{1}10]}_{TD}/(\tau b) = (\Delta L_{MD}^{max[\bar{1}10]}/\epsilon^{[110]})/(C\Delta t)$. Results of this normalization are plotted in Fig. \ref{fig:HeatExp}(e) as a function of the optical power density P. Accounting for the cycle to cycle strain variation, the trends initially observed (Fig. \ref{fig:HeatExp}(c)) are more clear. The Zn-core TD glide mobility is strongly decreased with optical power showing a more than two-fold reduction over the range in P in which the S-core glide shows a relatively flat response. At low optical power, $\mu _{Zn}$ is more than 10$\times$ that of S-core TDs, $\mu _{S}$. The photoplastic effect leads to a strong reduction in $\mu _{Zn}$, such that at the highest P measured, the Zn and S-core dislocations exhibit similar glide mobilities. Should the trend continue, we would expect $\mu _{S}>\mu _{Zn}$ at even higher optical powers. This reversal is in fact observed in photoplastic measurements of dislocations on additional samples in binary light/dark experiments (See Sec. \ref{app-photoplastic}).

These results are consistent with the mechanism of photoplasticity predicted in Ref. \cite{genlik_origin_2025} and illustrated in Fig. \ref{fig:photoplastic}(b), where photo-excited electrons are trapped by positively charged Zn-core dislocations forming new bonds within the core, greatly reducing their mobilities. That theory quantitatively predicts changes in the energetic Peierls barrier for edge dislocation glide, where electron-doped Zn-core pure edge dislocations show 800\% increase in the Peierls barrier, while hole-doped S-core dislocations show a slight increase of 13\%. For mixed 30° dislocations, the electron-doped Zn-core case shows a Peierls barrier increase of 760\%, while the S-core dislocations show a small reduction in the glide barrier by 22\%. These predictions are in remarkable agreement with the TD glide measurements shown in Fig. \ref{fig:HeatExp}(e). 

\section{n-type doping tunable dislocation glide}\label{doping-plasticity}
In the single dislocation photoplastic experiments, described above, light induces both photoelectrons and holes, thus, despite the close agreement with theory, the glide mobility dependence on optical excitation cannot experimentally separate the impact of electron or hole trapping effects independently. Additionally, these measurements are performed on minimally relaxed epilayers. Next we examine the impact of electronic (donor doping) of ZnS on the strain-relaxation process. As described above, in the undoped state, Zn-core dislocations are faster than S-core, leading to anisotropic strain relaxation\cite{FonsecaAlexandra2024}. With optical excitation, the Zn-core glide velocities are reduced, Fig. \ref{fig:photoplastic}(b). In the case of large n-type doping, we may expect a full reversal, Fig. \ref{fig:photoplastic}(c), with S-core glide velocities exceeding those of Zn-core, or the total prevention of Zn-core glide at all. To test these predictions, we use substitutional Al-doping to introduce electron doping in ZnS, see Methods.

\begin{figure} 
\centering
  \includegraphics[
  ]{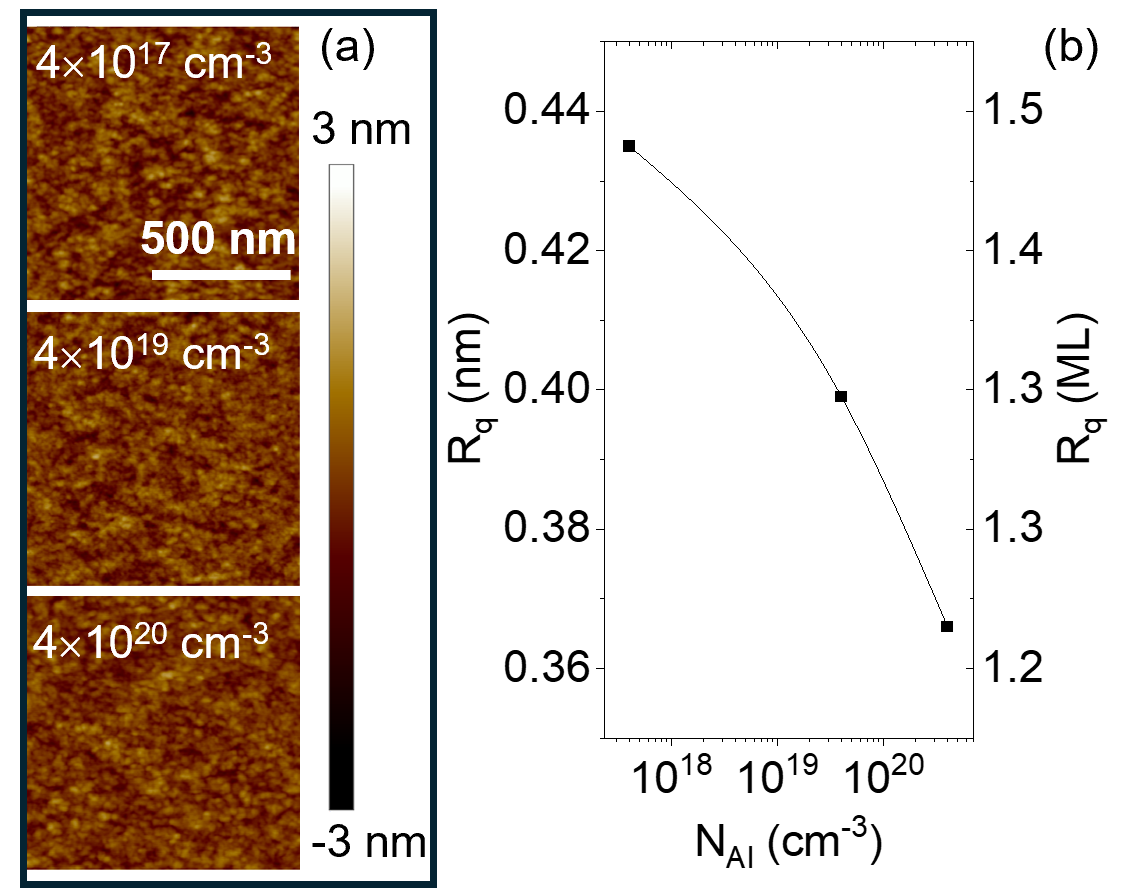}
  \caption{ZnS growth morphology as a function of n-type Al-doping concentration ($N_{Al}$). (a) AFM surface topology of 30-nm thick ZnS epilayers grown by Molecular beam epitaxy (MBE) on GaP (001). (b) RMS roughness ($R_q$) variation with $N_{Al}$. Line is a guide to the eye.}
\label{fig:AFM}
\end{figure}

It was previously found that the surface morphology of ZnS epilayers on GaP undergoes a transition from atomic layer-by-layer to 3D growth mode with thickness.\cite{FonsecaAlexandra2024} The transition coincides with a plasticity burst, a rapid increase in the density of  TDs. The resulting strain field at the surface could pin advancing atomic step edges leading to cumulative step-bunching. However, for the 30 nm Al-doped ZnS films, a smooth morphology is observed by atomic force microscopy (AFM) indicative of layer-by-layer growth, Fig. \ref{fig:AFM}(a) 
with an overall RMS roughness ($R_q$) <1.5 ML. The surface roughness shows a small but measurable decrease as the Al concentration ($N_{Al}$) increases, Fig. \ref{fig:AFM}(b). This suggests that with increasing $N_{Al}$, dislocations are being suppressed in the ZnS films. It was previously reported that n-type doping in a similar II-VI semiconductor, ZnO, at doping concentration of <0.1 at.\% resulted in films with lower defect density and decreased surface roughness \cite{SalahMohamed2018}.

\begin{figure*} 
  \includegraphics[width=\textwidth]{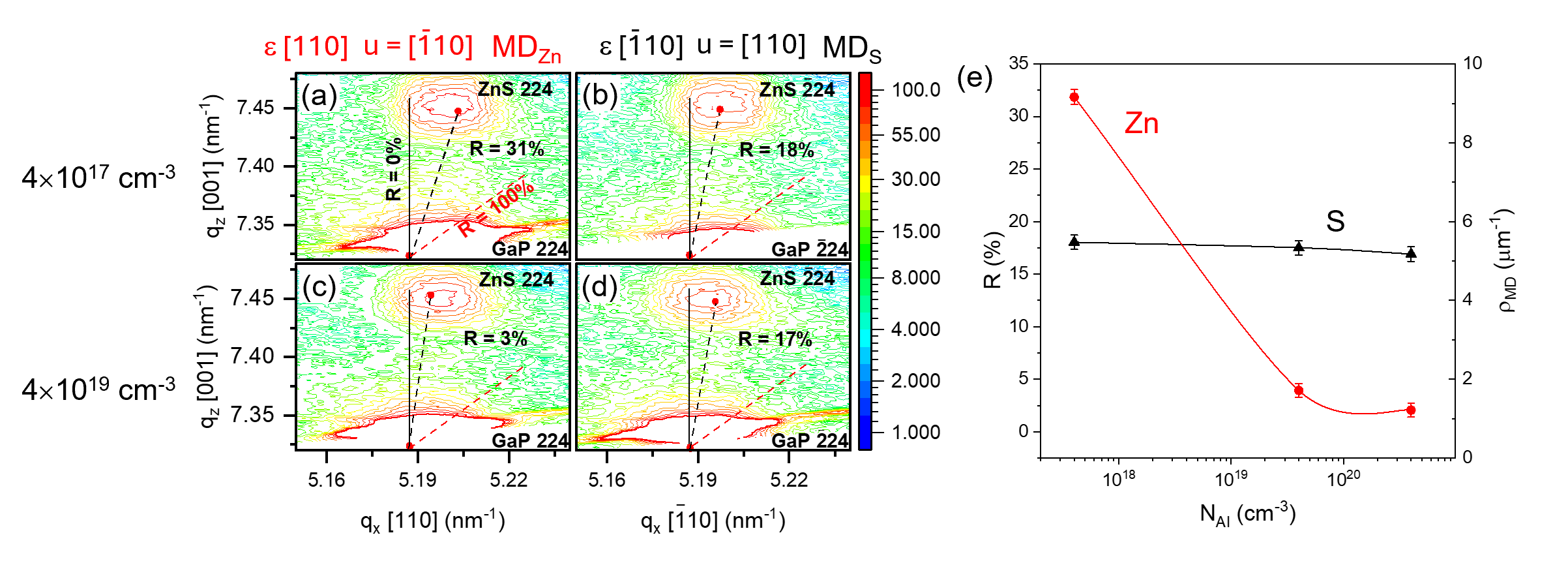}
  \caption{ZnS epitaxial strain-state as a function of n-type (Al) doping concentration ($N_{Al}$) measured by x-ray diffraction (XRD) reciprocal space mapping (RSM). (a) and (b) show the RSM at the (224) and ($\bar{2}24$) Bragg peaks, respectively, of a ZnS epitaxial layer with $N_{Al} = 4\times10^{17}$ cm$^{-3}$. Lines extending through the GaP Bragg peak represent various possible relaxation states of the ZnS layer. (c) and (d) show results of the same RSM scans for a ZnS sample with $N_{Al} = 4\times10^{19}$ cm$^{-3}$, showing a strong reduction in the in-plane strain $\epsilon$ along [110], with increased $N_{Al}$, but not along [$\bar{1}$10] indicative of an n-type doping suppression of Zn-core MDs (u=[$\bar{1}$10]), see Fig. \ref{fig:photoplastic}(c). (e) The corresponding MD densities $\rho_{MD}$ calculated from the RSM data as a function of $N_{Al}$ for Zn-core versus S-core MDs. Error bars are calculated by propagating the uncertainty in Gaussian peak fits of the RSM peak positions. Lines guide the eye.}
\label{fig:Relaxation}
\end{figure*}

To assess the influence of $N_{Al}$ on the dislocation density, x-ray diffraction reciprocal space mapping (RSM) is carried out to measure the in-plane strain-state, see Methods. \fig{Relaxation}(a) plots representative RSM data taken at the asymmetrical (224) reflection for a doped sample of $N_{Al} = 4\times10^{17}$ cm$^{-3}$ corresponding to 0.0008 at\%.
The vertical line intersecting the GaP (224) Bragg diffraction peak at $q_x [110] =2 \sqrt{2}a_{GaP}$, represents the in-plane lattice matching condition if the ZnS (224) diffraction peak lies along this line, such that along the [110] direction ZnS is fully strained, i.e. $(1-\epsilon^{[110]}/f) =R=0\%$ If the ZnS (224) peak lies along the diagonal red line, then the film would be fully relaxed, i.e. $R = 100\%$. The observed ZnS peak lies along the dashed black line indicating an $R = 31\%$, or $\epsilon^{[110]}/f=0.69$. From this value we determine the MD linear density with $\vec{u}$ in the orthogonal [$\bar{1}$10] direction, $\rho_{MD}^{[\bar{1}10]} = (2\sqrt{2}f/a)(1-\epsilon^{[110]}/f) = 9.1$ $\mu m^{-1}$ corresponding with Zn-core MDs formed by gliding Zn-core TDs. \fig{Relaxation}(b) shows the corresponding RSM for the same sample at the ($\bar{2}$24) diffraction peak showing a lower relaxation, $R = 18\%$, and correspondingly lower $\rho_{MD}^{[110]} = 5.3$ $\mu m^{-1}$ for the S-core dislocations. Increasing the n-type doping to $N_{Al} = 4\times10^{19}$ cm$^{-3}$ strongly suppresses strain-relaxation along [110], $R = 3\%$ and $\rho_{MD}^{[\bar{1}10]} = 0.9$ $\mu m^{-1}$ corresponding to a drastic suppression of Zn-core TD glide, Fig. \ref{fig:Relaxation}(c). In contrast, relaxation along [$\bar{1}$10] is relatively unchanged, $R = 17\%$ and $\rho_{MD}^{[110]} = 5.0$ $\mu m^{-1}$ showing the S-core TD glide to be insensitive to n-type doping, Fig. \ref{fig:Relaxation}(d). 

Based on RSM measurements of the n-type ZnS epilayers, the direction dependent relaxation, and corresponding dislocation densities are plotted in Fig. \ref{fig:Relaxation}(e) as a function of n-type doping. The effect of $N_{Al}$ is to inhibit Zn-core TD glide, resulting in a drop in the MD density by more than one order of magnitude. Since the the S-core density remains essentially unchanged with n-type doping, a crossover occurs such that above $10^{19}$ cm$^{-3}$ the S-core TD velocity surpasses that of the Zn-core TDs. These results mirror the optical power dependent effective TD mobilities plotted in Fig. \ref{fig:HeatExp}(e), where the $\mu _{Zn}$ strongly decreases under 365-nm optical illumination.  

\begin{figure}
\centering
    \includegraphics[
    ]{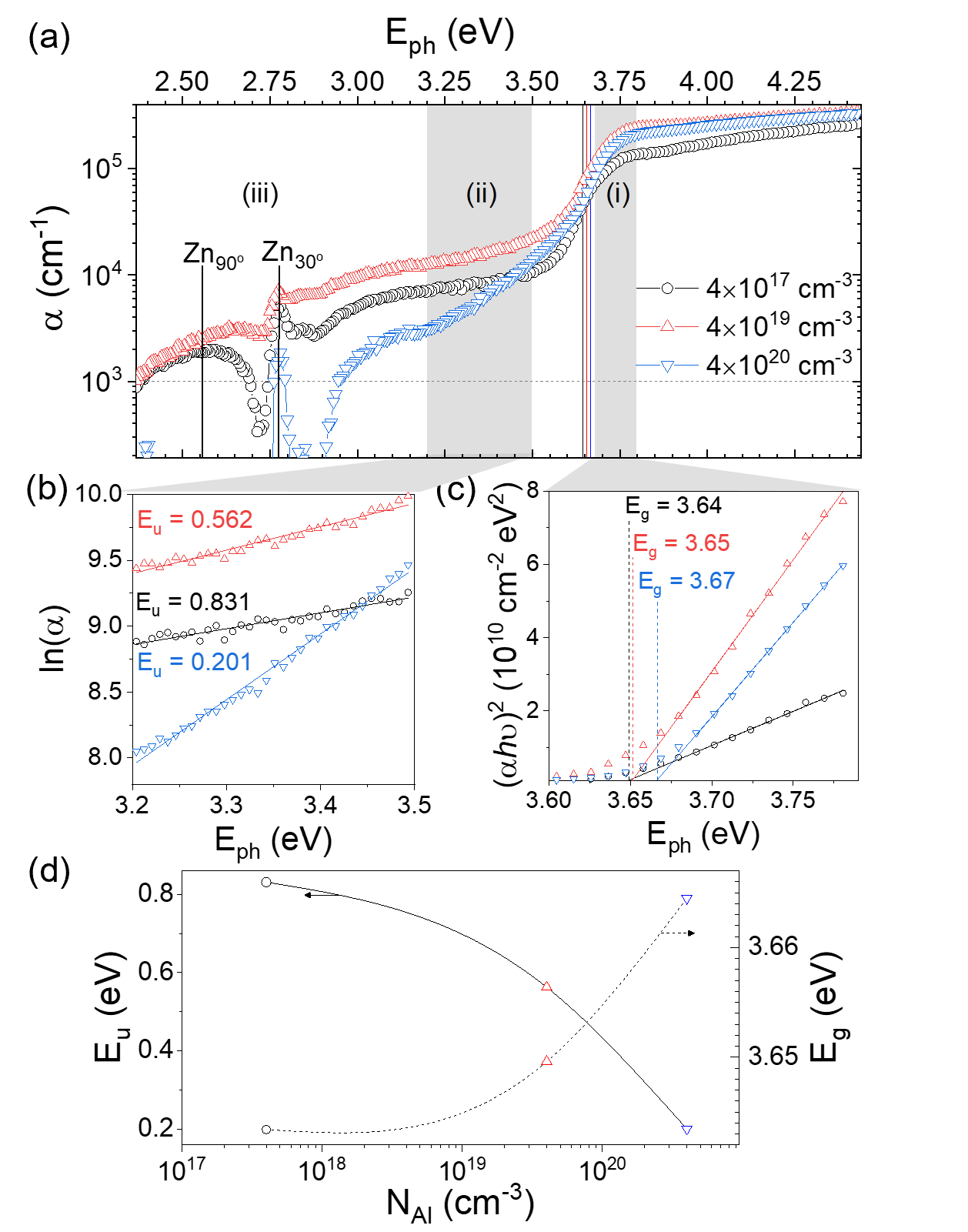}
    \caption{Impact of n-type (Al) doping on absorption spectra measured by spectroscopic ellipsometry. (a) Absorption spectra in ZnS epilayers grown with different Al doping concentrations (N\textsubscript{Al}). The three absorption regions (i)-(iii) described in the text, are labeled. Two absorption features in region (iii) are labeled according to the matching partial dislocation band gaps from Ref. \cite{genlik_origin_2025}.  (b) Urbach energy fit (lines) of the below band gap  band tail absorption. (c) Tauc fit (lines) of the absorption spectra above band gap. (d) Bandgap (E\textsubscript{g}) and Urbach energy (E\textsubscript{U}) as a function of N\textsubscript{Al} (lines are guides to the eye).}
    \label{fig:ellipsometry}
\end{figure}

Optical absorption spectroscopy is utilized to examine the impact of n-type doping on the electronic structure of ZnS. The absorption coefficient $\alpha$ as a function of photon energy $E_{ph}$ is plotted for the Al-doped ZnS epilayers in Fig. \ref{fig:ellipsometry}(a), based on spectroscopic ellipsometry, see Methods. The spectra illustrate three physically distinct 
absorption regions: (i) $E_{ph}\geq E_g$, interband (bulk conduction band to valence band) direct transitions, (ii) $E_{ph}< E_g$, disorder-induced (band tail) absorption, and (iii) $E_{ph}\ll E_g$,defect-band transitions (defect to bulk conduction or valence band). Signatures of (iii) are observed at 2.56 eV and a sharp peak at 2.78 eV, well below the band gap of ZnS. Both peaks reduce with increasing n-type Al-doping. These energies are close to the dislocation core related transitions predicted in Ref. \cite{genlik_origin_2025}, specifically for the Zn-core dislocation band gaps (2.27 eV and 2.94 eV, for $\theta$ = 90° and 30°, respectively). The reduction of these defect absorption transitions is coincident with the reduction in Zn-core MD density with n-type doping, which validate both the electronic structure (optical transitions) and Peierls barrier calculations of Ref. \cite{genlik_origin_2025}.

Crystalline defects introduce disorder in a semiconductor by interrupting the periodic potential, smearing the density of states below and above the conduction and valence band edges, respectively. These band tail states can be observed in the sub-band gap absorption spectra of semiconductors. Instead of a sharp cut-off in $\alpha$ for $E_{ph}<E_g$, a characteristic exponential decay is observed, the Urbach tail. The origin of this effect is not uniquely defined since it can arise in a number of different defect configurations (unlike the sharp defect band peaks of region (iii), however the Urbach tail is generically caused by a breakdown of the crystalline order\cite{dow_toward_1972,john_theory_1986,boubaker_physical_2011}. In region (ii) of the absorption spectra, such an Urbach tail is observed, and is fit to determine the Urbach energy $E_U$ as seen in Fig. \ref{fig:ellipsometry}(b), which shows a dramatic reduction as a function of n-type impurity doping, Fig. \ref{fig:ellipsometry}(d-e). This is consistent with the measured reduction in dislocation content with n-type doping implying that the Urbach tail in these partially relaxed epilayers is dominated by the strain-relaxation related dislocation content. Besides dislocations, impurity dopants also disrupt the periodic potential and induce an Urbach tail, however our results show that even at the highest Al doping levels, the Urbach tail is still decreasing, implying that it is dominated by the reduction in dislocation content. Region (i) of the spectra are Tauc plotted in Fig. \ref{fig:ellipsometry}(c) to determine the optical band gap of ZnS. A small blue shift of the band gap is observed as function of n-type doping, Fig. \ref{fig:ellipsometry}(d-e), which can arise from the finite filling of the conducton or valence band by doping. In our case, Al-doping leads to filling of the bottom of the ZnS conduction band, widening the gap, i.e. the Moss-Burstein shift\cite{lu_carrier_2007}.

Several experimental controls, described below, support that the measured differences arise from controlled electronic doping rather than uncontrolled structural variations. To minimize sample variation, all n-type ZnS films are grown on identical substrates within a single MBE session, with Al flux as the only variable. Ex-situ AFM, optical absorption, and HRXRD reciprocal space mapping (RSM) confirm that all samples possess comparable structural and optical quality. The Al-doped layers exhibit slightly smoother surfaces, and HRXRD peaks remain narrow and well-defined across all doping levels, indicating equivalent crystalline quality. The systematic variation appears only in the anisotropic strain relaxation, which scales with Al concentration, and in electronic signatures such as band filling and suppressed sub-bandgap absorption.

While potential contributions from solid solution strengthening can be evaluated independently, we note that in this case, the effect of Al on dislocations is primarily electronic in nature. Specifically, Al influences only the Zn-cores and not the S-cores. Since solid solution strengthening arises from structural mismatch between the solute and the host lattice, it would be expected to affect both types of dislocations. Moreover, the Al doping concentration used here is significantly lower than the typical solute concentrations—usually a few atomic percent—required to induce noticeable solid solution strengthening. These observations support the conclusion that the measured differences arise from controlled electronic doping rather than uncontrolled structural variations.

\section{Conclusions}\label{conclusions}
In this work, we have experimentally validated previous theoretical predictions regarding the photoplastic effect in ZnS. By directly imaging individual interface misfit dislocations under varying optical excitation conditions, we demonstrated that Zn-rich dislocation glide is drastically reduced with increasing optical power, while S-core dislocations show negligible sensitivity to light, in agreement with our computational predictions.  In addition, we targeted the behavior of Zn-core dislocations, by n-type doping of ZnS with Al. We found that, as the Al dopant density is increased, the resulting Zn-core dislocation density is reduced by more than one order of magnitude, while the S-core density remains
essentially unchanged. This is again in full agreement with the computational predictions as it shows that (a) positively charged Zn-core dislocations are affected by n-type doping, while the negatively charged S-cores are not affected, and (b) the excess electrons reduce the mobility of the Zn-rich dislocations and reverse the sign of the strain anisotropy. Although the effects of doping studied here are not pure examples of electroplasticity in the \textit{conventional sense} of current or E-field modulated mobility of dislocations, doping alters the charge carrier concentration and equilibrium Fermi level, which in turn modifies the charge state and mobility of dislocations. Thus, electroplasticity can still be explained by the same mechanism since under steady-state bias or current-injection the non-equilibrium, quasi-Fermi level, and charge state of dislocations is modulated. The observations in this study serve to demonstrate that the common origin of opto-electronic sensitivity of dislocation mobility in ZnS is the response of charged dislocations to excess carrier concentrations.

The present study was designed to test the theoretical predictions of Genlik et al. (2024) by examining how optical excitation and n-type doping affect dislocation glide and strain relaxation. The results are consistent with the predicted electronic mechanism, but do not by themselves constitute direct spectroscopic proof. Such spectroscopic measurement of optical transitions of dislocations in ZnS are the subject of a complementary study.\cite{blackston_photoluminescence_2025}

Beyond the immediate implications for ZnS, these findings contribute to a broader understanding of defect mediated plasticity in semiconductors. The correlation between optical absorption features and dislocation-induced localized states offers a non-destructive diagnostic tool for assessing defect density and electronic disorder. Moreover, the ability to modulate dislocation dynamics via light or doping opens avenues for dynamic control of mechanical properties in optoelectronic devices, potentially enabling adaptive materials that respond to external stimuli. Ultimately this research lays the groundwork for a new class of strain-engineered materials where electronic and mechanical properties are intricately intertwined and externally tunable.

\section{Methods}\label{Methods}
\subsection{Molecular Beam Epitaxy growth of ZnS on GaP}\label{Method-MBE}
ZnS epilayers are grown by molecular beam epitaxy on GaP (001) substrates at 30 nm thicknesses through methods described elsewhere with the addition of aluminum (Al) dopant.\cite{FonsecaAlexandra2024}. High-purity Al (99.999\%, Heeger Materials Inc.) are evaporated from an effusion cell at 1000, 1150 and 1300 °C. A beam equivalent pressure (BEP) measured from a nude ion-gauge in the substrate position gives a  flux of 7.39 × 10\textsuperscript{-9} Torr for T$_{Al}$= 1300 °C and undetermined for the temperatures below. Secondary Ion Mass Spectrometry (SIMS) analysis using Zn and Ga raw ion counts as layer markers, shows Al concentrations ($N_{Al}$) of 4 × 10\textsuperscript{17}, 4 × 10\textsuperscript{19} and 4 × 10\textsuperscript{20} cm\textsuperscript{-3} for 1000, 1150 and 1300 °C cell temperatures, respectively.

\subsection{Structural and Optical Characterization}\label{Method-Char}
Atomic force microscopy (AFM) on a Bruker Icon 3 AFM and high-resolution X-ray diffraction, including reciprocal space mapping (RSM) with a Bruker D8 HRXRD system, are employed to examine the surface roughness and layer relaxation. The relaxation measured through RSM is determined by $\omega$-2$\theta$ scans aligned to the symmetric (400), and the asymmetric (224) and ($\bar{2}$24) planes, where the diffraction plane is parallel to the [110] and the [$\bar{1}$10] crystallographic directions, respectively. For proper identification of the crystallographic directions, the flat directions given by the substrate's manufacturer are used to align the sample for HRXRD measurements. Optical properties are evaluated through spectroscopic ellipsometry (J.A. Wollam VASE 75 W) at isotropic conditions with 100 revolutions per measurement from 200 to 600 nm wavelengths with 1 nm steps at a fixed incident angle of 70°. The raw ellipsometry data are converted into n and k spectra by the Kramers–Kronig relations. The experimentally obtained parameters, $\psi$ and $\Delta$, determine the ratio of the Fresnel coefficients ($\rho$) through the equation \(\rho=r_p/r_s=tan(\psi)e^{i\Delta}\), where \(r_p\) and \(r_s\) are the parallel and perpendicular components of the light reflected from the interfaces, respectively \cite{rossow1996spectroscopic,Hilfiker2018}. From this ratio, the dielectric function is calculated as \(\varepsilon=sin^2(\theta_0)+sin^2(\theta_0) tan^2(\theta_0) [(1-\rho)/(1+\rho)]^2\), which is directly related to the optical constants by \(\varepsilon=\varepsilon_1+i\varepsilon_2=N^2=(n+ik)^2\), \(\theta_0\) being the incident angle \cite{rossow1996spectroscopic}. The absorption coefficient $\alpha$ is then calculated as \(\alpha=4k\pi/\lambda\) \cite{ROCHA2014436}. The Urbach energy is determined by fitting the Urbach absorption tail to the following equation \(\alpha \propto exp[(E-E_g)/E_u]\) \cite{KleinKampermann}. Finally, the optical bandgap is obtained through a linear fit of the Tauc plot on the band edge of the absorption spectra.

\subsection{Electron Contrast Channeling Immaging (ECCI) misfit dislocation photo-glide measurement}\label{Method-ECCI}
Undoped ZnS films grown on GaP (001), as described in previous studies \cite{FonsecaAlexandra2024}, at 20 nm thickness are temperature cycled in-situ in an environmental scanning transmission microscope equipped with a back-scatter electron detector (Thermo Scientific Quattro ESEM). The dislocations are imaged through electron channeling contrast imaging (ECCI) \cite{FonsecaAlexandra2024} at an accelerating voltage of 30 kV and a current of 2.4 nA at room temperature after each temperature cycle. The temperature cycling consists of annealing the films at 350 °C under different power intensities and cooling to room temperature for imaging. The low drift of the ESEM and heating stage enable imaging of the identical sample location and dislocations  before and after heat treatment. The light source used is a 365 nm light-emitting diode (LED) in which the power intensity at the sample position was varied as 0.67, 0.29, 0.09 and 0.07 mW/cm\textsuperscript{2}. A total of 8 micrographs were taken; one before annealing and one after annealing for each power intensity condition. Statistical analysis of the ECCI images was obtained from MIPAR image processing software \cite{Sosa_Huber_Welk_Fraser_2014,Blumer_Boyer_Deitz_Rodriguez_Grassman_2019}. In all, 148 misfit dislocations are quantified and analyzed. However, based on previous studies it was determined that dislocations are intrinsically prone to pinning, which hinders dislocation glide and motion \cite{FonsecaAlexandra2024}. Therefore, only the longest dislocations in an MD ensemble are representative of TD glide and consequently are the ones being considered for the effect of charge-carriers on dislocation motion.

Crystal directions are determined, as described in Sec.\ref{Method-Char}, by aligning the sample to the flat directions from the substrate. However, to confirm the crystal directions, the Kikuchi diffraction pattern, which relates to the electron diffraction pattern (ECP), is modeled using CrystOrient KLine software. It was determined that flat directions followed the stipulated crystal directions. Using the modeled ECP, the sample is tilted and rotated to align to the desired diffraction vector as detailed in earlier work \cite{FonsecaAlexandra2024}. 

\section{Acknowledgements} 
We gratefully acknowledge the support of this work  by the Air Force Office of Scientific Research grant FA9550-21-1-0278.

\section{Appendix} 
\subsection{Core structure and composition in misfit and threading dislocation segments of cubic ZnS}\label{App-dln}
\begin{figure} 
\centering
  \includegraphics[width=\columnwidth
  ]{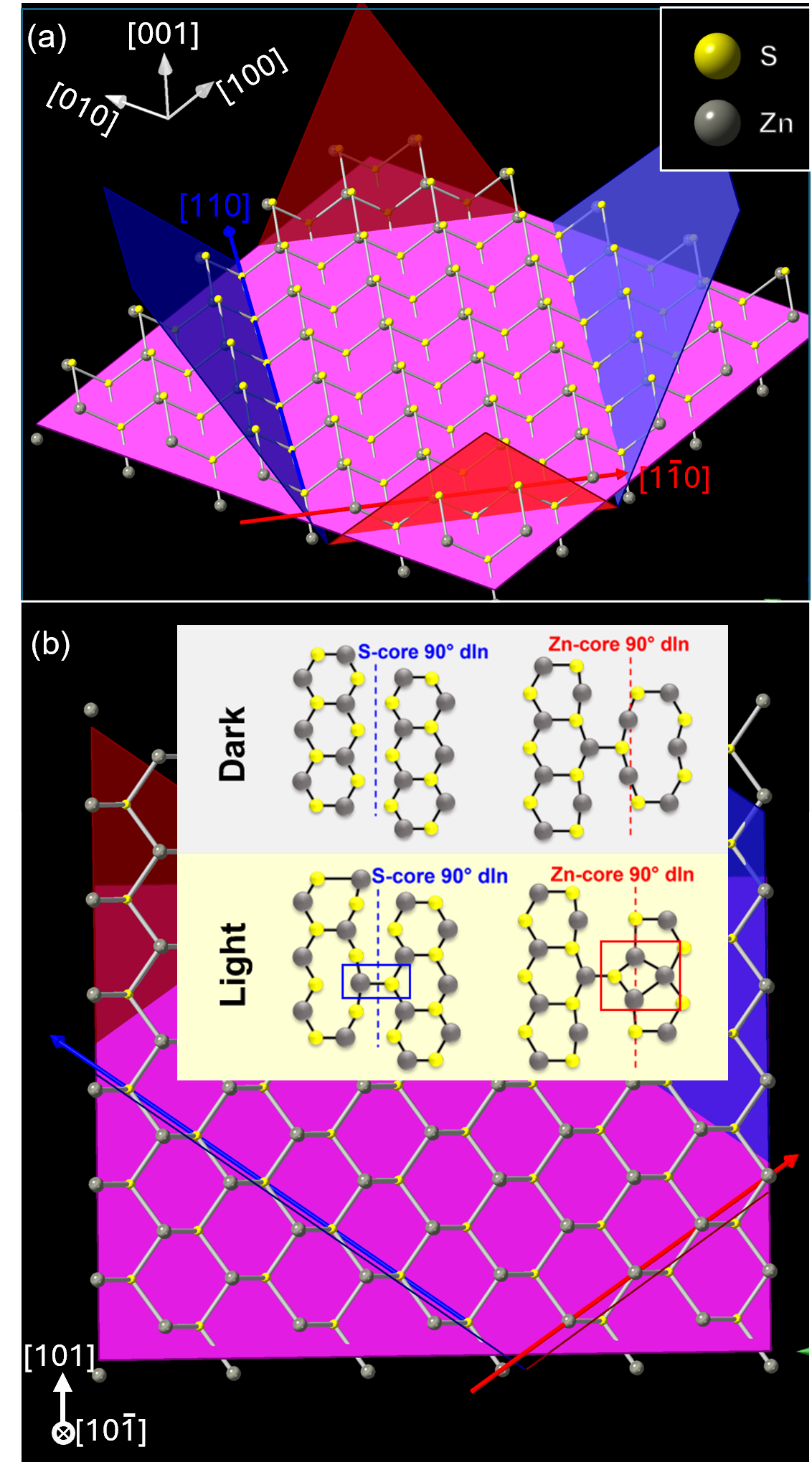}
  \caption{(a) The four possible \{111\} slip planes for dislocations in epitaxial zinc blende grown along [001]. The slip planes (red and blue) are coded to their surface termination (A or B). The (001) interface is colored magenta. Arrows indicate two possible MD line directions. (b) Parallel view of two of the slip planes of the glide-set illustrating the different terminating atom for two orthogonal <110> directions of MDs. Inset: atomic-scale structure of dislocation core structures of ZnS under light and dark conditions for the two compositions, adapted from Ref. \cite{genlik_origin_2025}.}
\label{slipplanes}
\end{figure}

For a (001) oriented substrate and epilayer, there are four \{111\} slip planes within which $60^{\circ}$ MDs may form in zinc blende epitaxial layers, see Fig. \ref{slipplanes}.
The substrate/epilayer interface is colored magenta. For a zinc blende crystal structure, an edge dislocation (or one with any edge character, $0^\circ<\theta<180^\circ$), gliding on a \{111\} plane, will contain an extra-half-plane (hp) terminating in either an A (e.g. Zn for ZnS) or B atom ( e.g. S) since the \{111\} planes in this crystal structure are either \{111\}A terminated (red) or \{111\}B terminated (blue), according to the unit cell convention that the [111] direction is defined for the A atom termination ([$\bar{1}\bar{1}\bar{1}$] for the B termination), e.g. Zn, terminating \{111\} plane. The orientation of the hp, either above or below the interface, depends on the sign of the epitaxial strain, $f < 0$ for compression (e.g. InGaAs/GaAs) and $f>0$ for tension (e.g. ZnS/GaP). Furthermore, depending on if the dislocations are glide or shuffle type slip planes, the terminating atom at the end of the hp will vary.\cite{petrenko_charged_1980, osipyan_properties_1986}. This combination of the sign of the misfit strain, glide/shuffle, and slip-plane results in a strict correspondence of the core composition of the MD with $\vec{u}$.\cite{haasen_plasticity_1957}. This is demonstrated in Fig. \ref{slipplanes} where orthogonal  $\vec{u}$ within a particular \{111\} slip-plane show opposite cation/anion termination. This geometric restriction was first pointed out in InSb by Haasen,\cite{haasen_plasticity_1957} and later generalized for all zinc blende structures by Abrahams, Blanc, and Buiochhi.\cite{abrahams_likesign_1972} Table \ref{coretable} illustrates all possible cases for the dislocation core composition in zinc blende epitaxial layers on (001) oriented substrates. For the case of ZnS in tension, edge component dislocations (MDs) with $\vec{u}=[110]$ contain S-core and $\vec{u}=[\bar{1}10]$ contain Zn-core. The literature frequently reference dislocation types according to $\alpha/\beta$ labeling, where $\alpha$ ($\beta$) dislocations terminate in A (B) atoms for the shuffle-set, yet this leads to confusion for materials in which the glide-set are active. To avoid confusion, the Hünfeld convention is adopted,\cite{alexander_international_1979} which specifies both the core termination and type, see Table \ref{coretable}.
\begin{table}[b]
\caption{\label{coretable}%
Relation of core composition and line direction for MDs in (001) oriented zinc blende epitaxial layers for all combinations of epitaxial strain, slip planes, and glide or shuffle-set. The core structure is labeled according to the Hünfeld convention, X(s/g), where X is the atom terminating at the half-plane (the core enriching atom), and the (s/g) whether it is shuffle or glide type. The obsolete $\alpha/\beta$ labels are included for completeness.}

\begin{ruledtabular}
\begin{tabular}{lccccr}
Set & $f$ & $\vec{u}$ & slip-plane & X(s/g) & $\alpha/\beta$ \\
\colrule
\colrule

 \multirow{4}{*}{shuffle} & \multirow{2}{*}{-} & [110] & (1$\bar{1}$1)B,($\bar{1}$11)B & B(s) & $\beta$ \\
                          &                    & [1$\bar{1}$0] & (111)A,($\bar{1}$$\bar{1}$1)A & A(s) & $\alpha$ \\ \cline{2-6}
                          & \multirow{2}{*}{+} & [110] & (1$\bar{1}$1)B,($\bar{1}$11)B & A(s) & $\alpha$ \\                      
                          &                    & [1$\bar{1}$0] & (111)A,($\bar{1}$$\bar{1}$1)A & B(s) & $\beta$ \\ \hline
 \multirow{4}{*}{glide} & \multirow{2}{*}{-} & [110] & (1$\bar{1}$1)B,($\bar{1}$11)B & A(g) & $\beta$ \\
                          &                    & [1$\bar{1}$0] & (111)A,($\bar{1}$$\bar{1}$1)A & B(g) & $\alpha$ \\ \cline{2-6}
                          & \multirow{2}{*}{+} & [110] & (1$\bar{1}$1)B,($\bar{1}$11)B & B(g) & $\alpha$ \\                      
                          &                    & [1$\bar{1}$0] & (111)A,($\bar{1}$$\bar{1}$1)A & A(g) & $\beta$ \\

\end{tabular}
\end{ruledtabular}
\end{table}
\begin{table*}[b]
\caption{\label{dlntable}%
Line directions and slip planes for all eight possible burgers vectors of strain relieving MDs and their associated TD pairs in zinc blende (001) epilayers under tensile strain. For ZnS (glide-set) the core composition and charge are listed to the right of the $\theta$ of each dislocation segment. The subscript (L/R) refers to the helicity of the screw-component of the TD, either left or right handed.}
\begin{ruledtabular}
\begin{tabular}{lcccccccccr}
$\vec{u}_{MD}$ & slip-plane & $\vec{u}_{TD_L}$ & $\vec{u}_{TD_R}$ & $\vec{b}$ & \multicolumn{2}{c}{$\theta_{TD_L}$} & \multicolumn{2}{c}{$\theta_{MD}$} & \multicolumn{2}{c}{$\theta_{TD_R}$}\\
\colrule
\colrule
\multirow{4}{*}{[110]} &  \multirow{2}{*}{ (1$\bar{1}$$\bar{1}$)} &   \multirow{2}{*}{[101]} &  \multirow{2}{*}{[01$\bar{1}$]}  & [01$\bar{1}$] & $120^{\circ}$ & S$^-$ & $60^{\circ}$ & S$^-$ & $0^{\circ}$& ZnS$^0$ \\ \cline{5-11} 
                                                                                                       &&&& [$\bar{1}$0$\bar{1}$] & $180^{\circ}$ &ZnS$^0$ & $120^{\circ}$& S$^-$ & $60^{\circ}$ &S$^-$\\ \cline{2-11}
  & \multirow{2}{*}{ (1$\bar{1}$1)} & \multirow{2}{*}{[10$\bar{1}$]} &   \multirow{2}{*}{[011]}  & [011] & $120^{\circ}$ &S$^-$ & $60^{\circ}$ &S$^-$ & $0^{\circ}$&ZnS$^0$ \\ \cline{5-11} 
                                                                                                       &&&& [$\bar{1}$01] & $180^{\circ}$&ZnS$^0$ & $120^{\circ}$&S$^-$ & $60^{\circ}$&S$^-$\\ \hline
 \multirow{4}{*}{[$\bar{1}$10]} &  \multirow{2}{*}{ (111)} & \multirow{2}{*}{[01$\bar{1}$]} &  \multirow{2}{*}{[$\bar{1}$01]}  & [$\bar{1}$01] & $120^{\circ}$&Zn$^+$ & $60^{\circ}$&Zn$^+$ & $0^{\circ}$&ZnS$^0$ \\ \cline{5-11} 
                                                                                                       &&&& [0$\bar{1}$1] & $180^{\circ}$ &ZnS$^0$ & $120^{\circ}$ &Zn$^+$ & $60^{\circ}$& Zn$^+$\\ \cline{2-11}
  &  \multirow{2}{*}{ ($\bar{1}$$\bar{1}$1)} & \multirow{2}{*}{[011]} &  \multirow{2}{*}{[$\bar{1}$0$\bar{1}$]}  & [$\bar{1}$0$\bar{1}$] & $120^{\circ}$& Zn$^+$ & $60^{\circ}$ &Zn$^+$ & $0^{\circ}$&ZnS$^0$ \\\cline{5-11} 
                                                                                                       &&&& [0$\bar{1}$$\bar{1}$] & $180^{\circ}$&ZnS$^0$ & $120^{\circ}$&Zn$^+$ & $60^{\circ}$&Zn$^+$\\ 
\end{tabular}
\end{ruledtabular}
\end{table*}
As discussed in the manuscript, the nucleation and elongation of MDs occurs due to the glide of a pair of TDs, which in turn share the same burgers vector, $\vec{b}$. For each of the four slip planes, two $\vec{b}$ are possible for $60^\circ$ or $120^\circ$ dislocations, thus there are eight total MDs, each with a pair of TDs. Table \ref{dlntable} presents the various dislocations possible for (001) oriented epitaxial ZnS containing strain-relieving MDs for $f>0$ (tensile), i.e. with hp oriented above the (001) interface. This geometry results in each MD with one of the TDs having identical edge-component and core composition (Zn or S-rich), yet opposite helicity (sign of screw component, $b_{screw}=\vec{b}cos(\theta)$) as that of the MD. The secondary TD is always a pure screw ($0^{\circ}$ or $180^{\circ}$) with a neutral ZnS composition, but having identical helicity as the screw component of its MD.

\subsection{Photoplasticity: reproducibility and statistics}\label{app-photoplastic}
\begin{figure*}
\centering
  \includegraphics[width=\textwidth
  ]{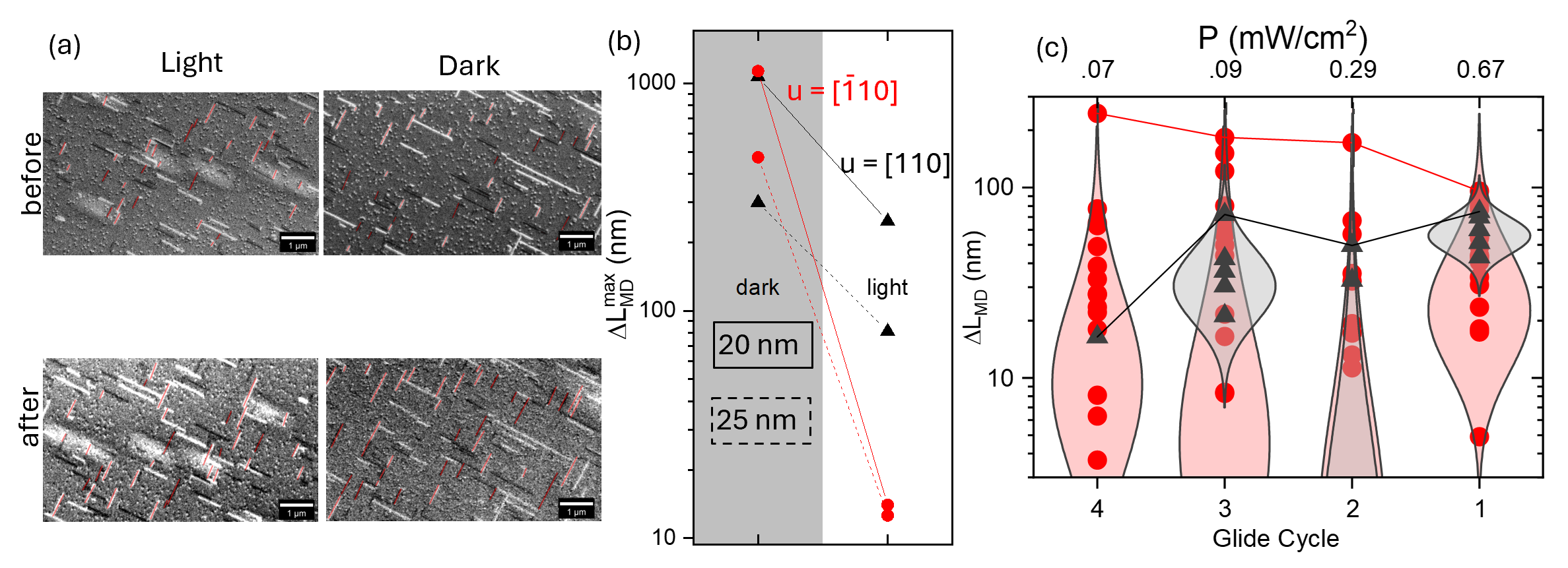}
  \caption{(a) Representative ECCI images with MDs along [$\bar{1}$10] quantified using MIPAR. The images are before and after a glide cycle under different lighting condition to quantify the photoplastic effect from individual MD length variation. (b) Maximum MD length change for two orthogonal line directions (u) extracted from ECCI, as in (a), showing the anisotropic impact of optical exposure. Measurements are on two different thicknesses of epitaxial ZnS on GaP. (c) Optical power dependent MD length changes, e.g. Fig. \ref{fig:HeatExp} of the main text. Violin plots of log-normal distribution fits are shown representing all MD length change measurements per glide cycle. Tie lines connect the maximum MD length change in each image. Vertical axis is plotted on a semi-log scale.}
\label{photo-ECCI}
\end{figure*}

The ECCI-based photoplastic experiment (Sec. \ref{single-photoplastic}) is carried out on additional ZnS samples to check the reproducibility of the effect. The same glide cycle procedure (Sec. \ref{Method-ECCI}) is carried out but under only two optical conditions, dark or light. Example raw SEM ECCI images of MDs are plotted in Fig. \ref{photo-ECCI}(a)
together with the MIPAR segmented MDs (red lines). For identical ECCI condition and image location, the MD content is quantified before and after glide. The maximum MD length change is plotted in Fig. \ref{photo-ECCI}(b) on a semi-log scale for dark versus light conditions.  The 20 nm film shows a two order of magnitude reduction in TD glide under illumination for the Zn-core [$\bar{1}$10] MDs, while the S-core [110] MDs show a smaller $5\times$ reduction. In the thicker film (25 nm) the same trend is observed but with reduced magnitude, the Zn-core [$\bar{1}$10] reduction is one order of magnitude, while that of the  S-core [110] is $2\times$. This is because the thicker film (25 nm) has a higher concentrations of dislocations, which is likely limiting the glide through pinning. However, in both films the binary (light/dark) glide reductions are larger than in the systematic P dependent study with the effect of light being a full reversal of the glide anisotropy, such that in darkness, Zn-core glide is larger than S-core, but in light the order is reversed.

To examine the statistical distribution of photo-modulated TD glide, we analyze the full distribution of MD length changes in the experiment described in Sec. \ref{single-photoplastic}. Figure \ref{photo-ECCI}(c) plots histograms of $\Delta L_{MD}$ fitted by log-normal distributions for each optical power density (P). The vertical axis is log-scaled, such that the distributions appears normal. As previously mentioned, the majority of MD length changes are limited by pinning. The maximum $\Delta L_{MD}$ in each image-set reveals the impact of optical illumination on unpinned TDs. For [$\bar{1}$10] Zn-core, $\Delta L_{MD}^{max}$ shows a monotonic reduction with optical power. S-core [110] show a non-monotonic trend. At the highest P, the Zn-core glide is reduced nearly to that of the S-core.

\bibliography{references}
\end{document}